\documentclass[aps,prb,twocolumn,10pt]{revtex4-2} 

\usepackage{amssymb,amsmath}
\usepackage{graphicx}
\usepackage{natbib}
\usepackage{multirow}

\usepackage[utf8]{inputenc}

\begin{document}

\title{Special behavior of alkali beam emission spectroscopy in low-ion-temperature plasma}

\author{P. Balázs$^{1,2}$, O. Asztalos$^{1,2}$, G. Anda$^{2}$, M. Vecsei$^{2}$, S. Zoletnik$^{2}$, S. T. A. Kumar$^{3}$, G. I. Pokol$^{1,2}$}

\affiliation{$^1$Institute of Nuclear Techniques, Budapest University of Technology and Economics, Budapest, Hungary}
\affiliation{$^2$Centre for Energy Research, Budapest, Hungary}
\affiliation{$^3$Department of Electrical and Computer Engineering, HSX Plasma Laboratory, University of Wisconsin-Madison, United States of America}

\begin{abstract}
Beam emission spectroscopy (BES) is a powerful plasma diagnostic method especially suited for the measurement of plasma density and its fluctuations. As such, synthetic BES codes are regularly used to aid the design or utilization of these diagnostic systems. However, synthetic diagnostics can also be used to study the method in previously not yet explored operational conditions. This paper presents such an analysis utilizing the RENATE\nobreakdash-OD synthetic diagnostic code for a hypothetical alkali BES system on the HSX stellarator. HSX is a device featuring an unusual operating regime in the world of fusion devices due to the low ion temperature and low plasma density. It was found that BES shows unusual tendencies in these conditions. The relation between beam energy and plasma penetration in low-ion-temperature plasma, together with unique emission features facilitated by low-density plasma, and the underlying reasons behind these features are explored in this paper.
\end{abstract}

\maketitle

\section{Introduction}

In the current development stage of fusion technology, plasma diagnostic tools are still much needed to support the validation of various theories and verification of engineering solutions. Active diagnostic techniques, when some kind of probing of the plasma is carried out, are especially attractive due to their localized measurements. Such a system is beam emission spectroscopy (BES), used in several experimental devices, including JET \cite{JET_Brix_2012}, ASDEX Upgrade \cite{Willensdorfer_PPCF_2014}, EAST \cite{EAST_BES_2017}, DIII-D \cite{DIII-D_ABES_1995}, W7-X \cite{anda2019measurement}, and even small devices like COMPASS \cite{Compass_BES_2016}. This technique probes the plasma with a neutral atomic beam, resulting in the release of light due to interactions with the plasma particles. The wavelength of this light is characteristic to the beam material, which can be one of the hydrogen isotopes in larger beams, or alkali metals (lithium, sodium) in the small diagnostic beams. The intensity of the emitted light is highly dependent on the plasma density, which combined with the ability of continuous operation makes the technique ideal for spatially and temporally resolved density fluctuation measurements. This is especially true for alkali diagnostic beams, with the added benefit of the emission spectrum being well distinguishable from the background. The limitation of this type of BES is the generally short, few centimeters long penetration into the plasma, but the region spanned by this limitation is still interesting in terms of transport and magnetohydrodynamic phenomena.

The construction of a BES system requires careful design due to geometric restrictions and the desire to maximize the collectable photon flux \cite{Zoletnik_RSI_ABES_2018}. As an aid for this problem, one can utilize BES simulation codes designed to produce synthetic measurements under the specific plasma conditions of the device in question. Based on the data of such simulations it is possible to analyze the expected performance of the diagnostic system, helping the engineers to balance between design trade-offs. RENATE\nobreakdash-OD \cite{rod_github} (Rate Equations for Neutral Alkali-beam TEchnique - Open Development) is an open-source Python package developed for this purpose. This code solves rate equations to calculate the population of each atomic level considered in the calculations, from which the emissivity of the beam is acquired through the spontaneous transition rate for the observed spectral line. With the ability to accurately predict the performance of a BES system under design, it is possible to dynamically test various injection and observation configurations for optimal operation.

RENATE\nobreakdash-OD and its precursor, RENATE \cite{pusztai2009deconvolution}, has already been utilized in multiple projects regarding BES systems. The code's capabilities have been validated with KSTAR measurements \cite{nam2012analysis}, it was the main tool for a feasibility study of a BES system for JT60-SA \cite{Asztalos_FED_2017}, and aided the design of lithium beam diagnostics on EAST \cite{Zoletnik_EAST_RSI_2018}. Details of how beam evolution is calculated by RENATE and RENATE\nobreakdash-OD can be found in \cite{Guszejnov_RSI_2012}. Recently, we performed a feasibility study for the HSX (Helically Symmetric eXperiment) stellarator. This device is operated by the Electrical and Computer Engineering department of the University of Wisconsin-Madison, with the aim of investigating transport, turbulence, and confinement in a quasi-helically symmetric magnetic field \cite{anderson1995helically}. It is a small device with average major and minor radii of 1.2 m and 0.15 m, respectively. The magnetic field is produced by copper coils, allowing discharges up to 100 ms long, and a maximum on-axis magnetic field of 1.25 T. The plasma density achievable in the device is on the order of a few times 10$^{18}$ m$^{-3}$, which is relatively low compared to other fusion-related plasma experiments. As a consequence, the electrons heated by electron cyclotron resonance are poorly coupled to the ions, resulting in an electron and ion population with disparate temperatures. Throughout this paper, we treated the ion temperature as constant at 50 eV throughout the machine, while the electron temperature was prescribed with a peaked profile reaching a maximum of 2.5 keV \citep{guttenfelder2008effect}.

As of now, the device does not have a BES system, therefore we performed a feasibility study to explore how the technique would perform under such conditions. This is particularly interesting since turbulent phenomena in stellarators are different from those in tokamaks, yet BES measurements have been rare so far in stellarators. W7-X has a lithium BES system installed \cite{anda2019measurement}, but being a high-plasma-density device, the measurements are limited to scrape-off layer and island divertor regions, while the low density of an HSX-like plasma promises greater penetration.

The study was based on the simulation of lithium (Li) and sodium (Na) beams across HSX-like low-density and low-ion-temperature plasma profiles. We observed that both the beam density and emissivity can behave unusually under these circumstances. Notably, a beam with low energy could be less attenuated than a high-energy beam, and changes in the electron temperature are more visible in the emission compared to measurements in high-density plasma.

The particular choices in the design of a BES system, for example, beam material and beam energy, can only be made with the help of a detailed study taking the possible observation geometries into account, while also considering penetration depth, spatial resolution, and signal-to-noise ratio (SNR) requirements. While the simulations presented here would be the first steps to inform such a study of a BES system for HSX, in this work we are solely focusing on the unique phenomena observable in a low-density, low-ion-temperature plasma, and the reasons behind these effects. We also note that these plasma conditions are not exclusive to HSX, since they can be regularly found in the divertor region of larger fusion devices. Measurement of these regions by BES is an unexplored area, which warrants further efforts for investigation. The basic methodology of the calculations performed by RENATE\nobreakdash-OD are presented in Chapter \ref{calc}, then in Chapter \ref{res} the unusual effects found in beam attenuation and emission are presented together with their explanation. Finally, in Chapter \ref{sum} the findings and their relevance are summarized.

\section{Calculations}
\label{calc}

To simulate the density and emission of an atomic beam along its path, RENATE\nobreakdash-OD solves the rate equations describing the population of the valence electrons of the beam atoms on the most populated atomic levels. In the case of alkali atoms, the considered levels are the l-resolved states reachable by the valence electron, with an energy up to 4.5 eV in the case of lithium, which includes 2s, 2p, 3s, 3p, 3d, 4s, 4p, 4d, 4f, and up to 4.1 eV for sodium, including 3s, 3p, 3d, 4s, 4p, 4d, 4f, 5s. The populations of the levels evolve due to interactions with the plasma components. The simulated light emission is governed by collisional excitation to the higher energy levels followed by spontaneous emission. The attenuation of the beam is contributed to the beam atoms becoming charged, and therefore redirected by the perpendicular magnetic field component. This can happen through collisional ionization or charge exchange with plasma ions, and we refer to the sum of these processes as electron loss. The simulation does not include interaction between beam atoms due to being negligible compared to beam-plasma interactions. It is also assumed that only the valence electrons participate in the relevant processes, and initially, all of them are in the ground state.

The atomic levels can gain or lose electrons through multiple channels. For example, if we denote the population density of a particular level with $n_i$ in a beam with given material and energy, we can write the losses of this level due to excitation by electrons in the form of

\begin{equation}
\left(\frac{dn_i}{dt}\right)_{el,exc}=-n_in_e\sum_{i<j}R_{i\rightarrow j}^{e,exc}(T)\ ,
\label{equ:dni}
\end{equation}

where $n_e$ is the electron density in the plasma, $R_{i\rightarrow j}^{e,exc}(T)$ is the rate coefficient expressing the probability of electron excitation from level $i$ to $j$, and $T$ is plasma temperature. In this form, it is assumed that the plasma has a Maxwell velocity distribution, with an insignificant bulk velocity compared to the beam velocity. The temperature dependence of the rate coefficient originates from the way it is calculated with the integral

\begin{equation}
R=\langle\sigma v\rangle=\int \sigma(v)vf(v,T)dv\ ,
\label{equ:rate}
\end{equation}

where $R$ is a general rate coefficient, $\sigma(v)$ is the cross-section of the reaction, $v$ is the relative velocity of the beam atom and plasma particle, and $f(v, T)$ is the velocity distribution of the plasma particles. The cross-sections for the different reactions are based on \cite{wutte1997cross, schweinzer1999database, schweinzer1994study} for lithium, and on \cite{igenbergs2008database} for sodium.

If we collect all the terms similar to the electron excitation term in equation (\ref{equ:dni}), namely excitation, de-excitation, and electron loss caused by all plasma components, we get an equation for the temporal evolution of $n_i$. However, we would like to express the properties of the beam spatially along its path ($x$). Assuming that the beam atoms travel with a constant speed ($v_b$), the conversion is simply

\begin{equation}
\frac{dn_i}{dx}=\frac{1}{v_b}\frac{dn_i}{dt} .
\label{equ:dxdt}
\end{equation}

In this sense, the spatial evolution of a beam can be expressed with the reduced rate coefficients, which are the regular rate coefficients divided by the beam velocity:

\begin{equation}
\bar R=\frac{R}{v_b}\ .
\label{equ:redrate}
\end{equation}

This is the main quantity we study closely when we want to explain specific phenomena in beam evolution. Its values were precalculated for each type of interaction for plasma temperatures in the range of $1-10^5$ eV, so during the simulation of beams $\bar R$ is looked up according to the local plasma temperature for each point of the beam.

\section{Results}
\label{res}

\begin{figure*}
\begin{center}
\includegraphics[width=.4\linewidth]{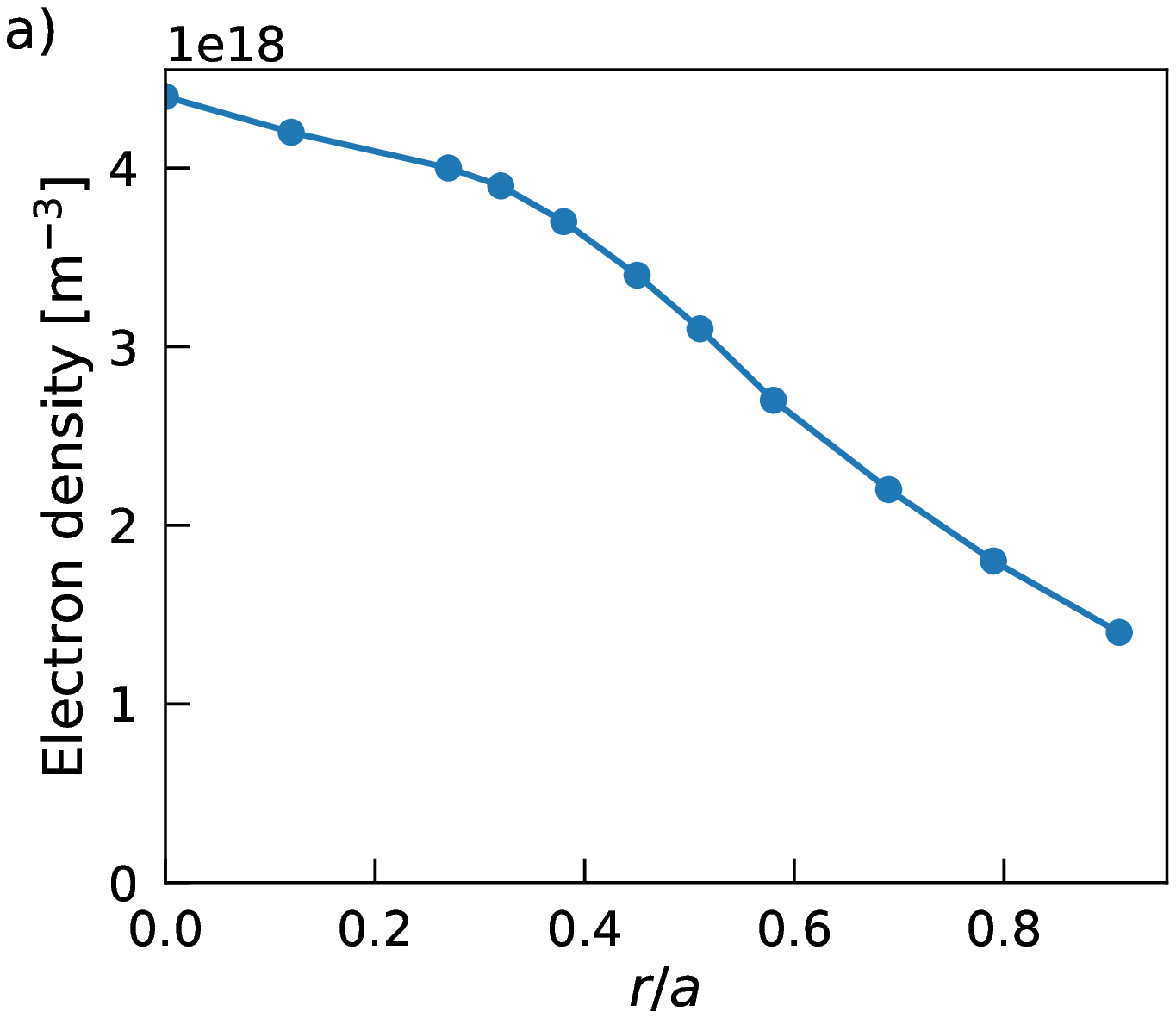}
\includegraphics[width=.4\linewidth]{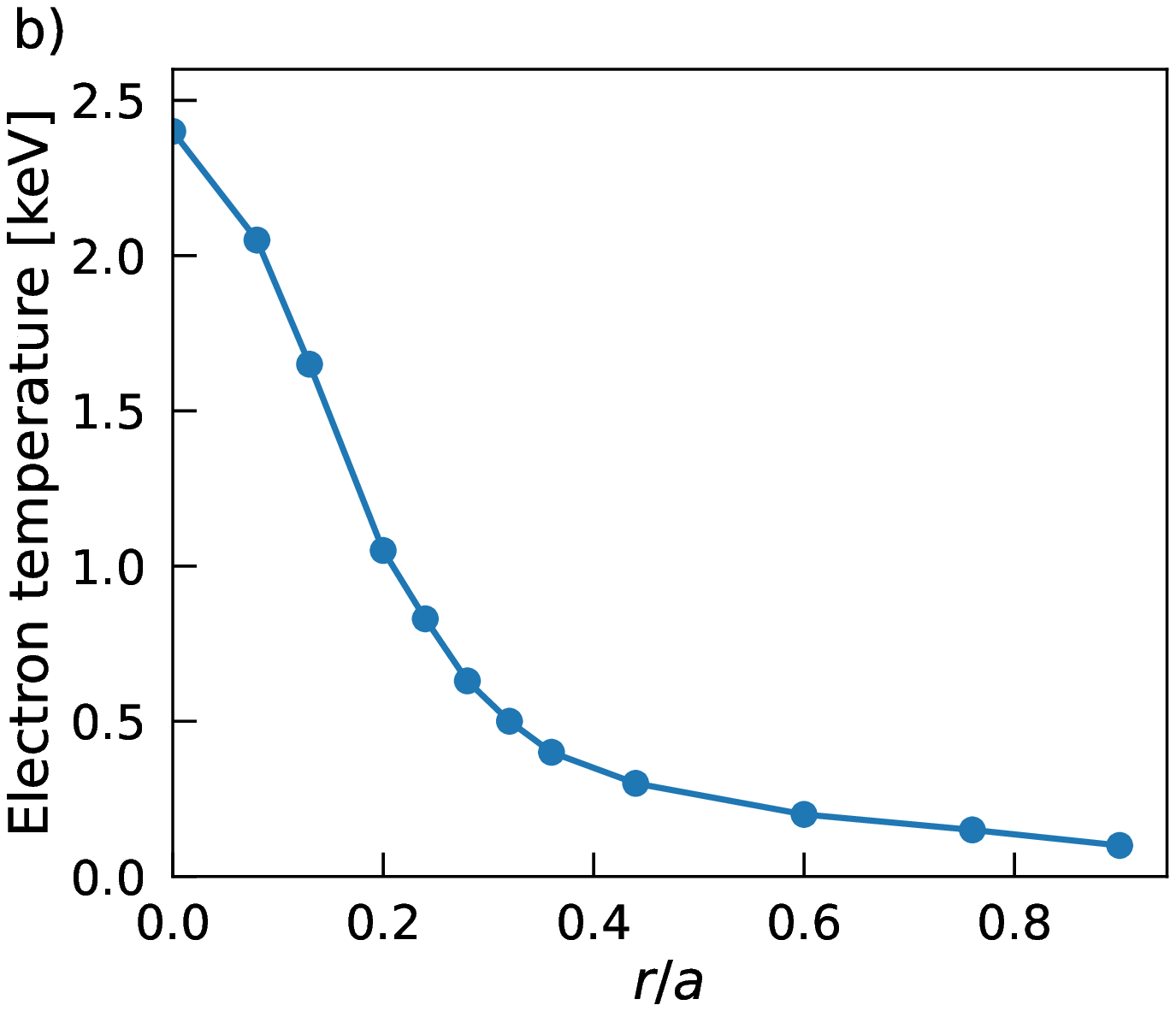}
\caption{Electron density (a) and temperature (b) profiles based on \cite{guttenfelder2008effect}.}
\label{fig:dens_T}
\end{center}
\end{figure*}

In order to perform the necessary beam simulations for the HSX study, the plasma composition together with the density and temperature profiles were required as inputs. We assumed a pure hydrogen plasma as a simple test case, which also meant equal proton and electron density in accordance with charge neutrality. The profiles were based on previous publications \cite{guttenfelder2008effect} on the results of the experiment. Figure \ref{fig:dens_T} shows the radial electron density and temperature profiles, with the minor radius ($r$) normalized to the device's minor radius ($a$). The experimental data only extended to about $r/a=0.9$. However, the density outside of this region is low enough to safely ignore in the simulations. This approach eliminates the need to perform any ad hoc extrapolation, effectively assuming zero density outside the simulated region. For ions, the same density profile was used, and the ion temperature was assumed to be homogeneously 50 eV according to \cite{kumar2017determination}. This is different from the conditions found in other devices because usually, a higher plasma density ensures that the electron and ion population has a similar temperature.

\subsection{Beam density}

\begin{figure*}
\begin{center}
\includegraphics[width=.45\linewidth]{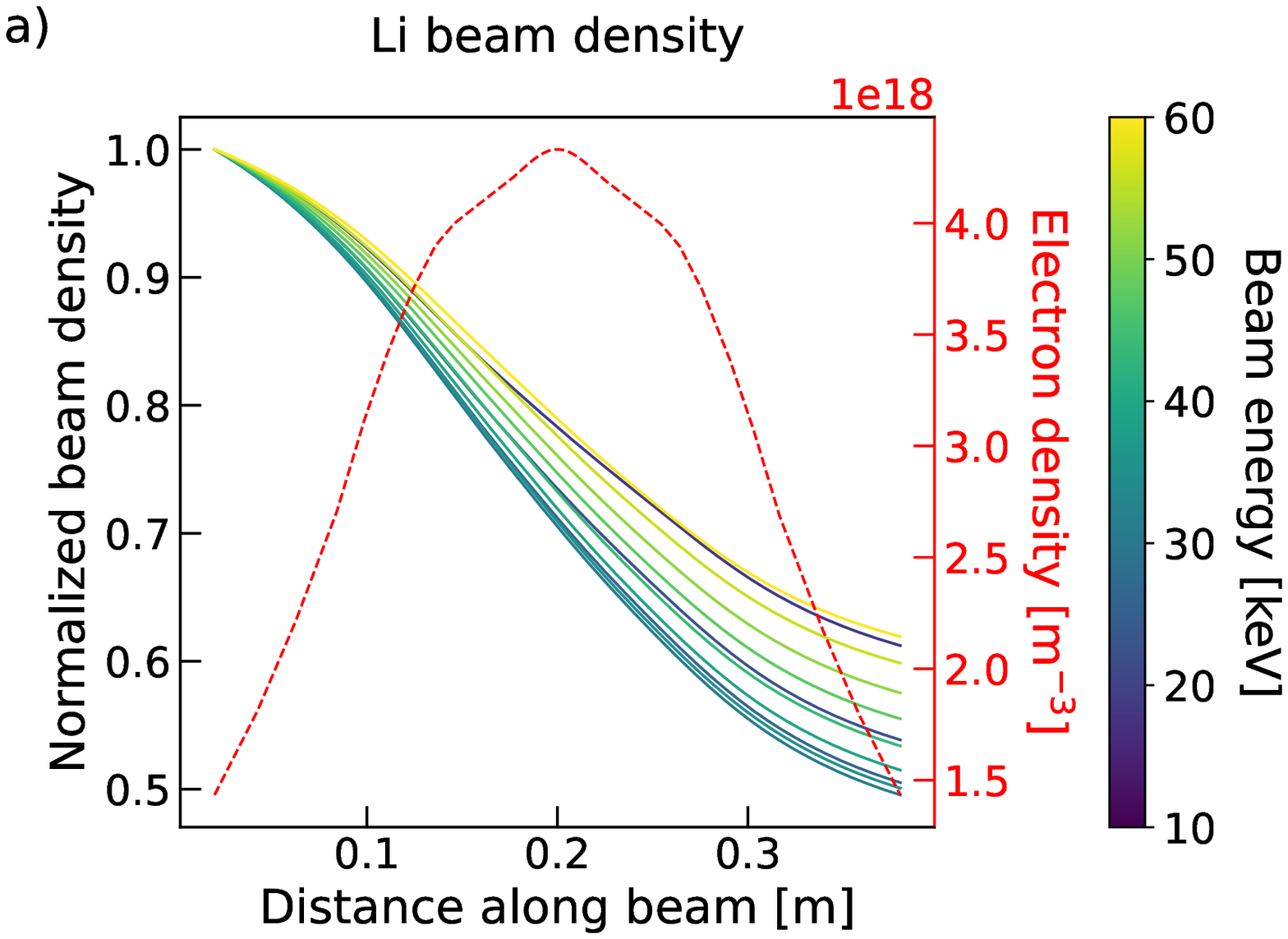}
\includegraphics[width=.45\linewidth]{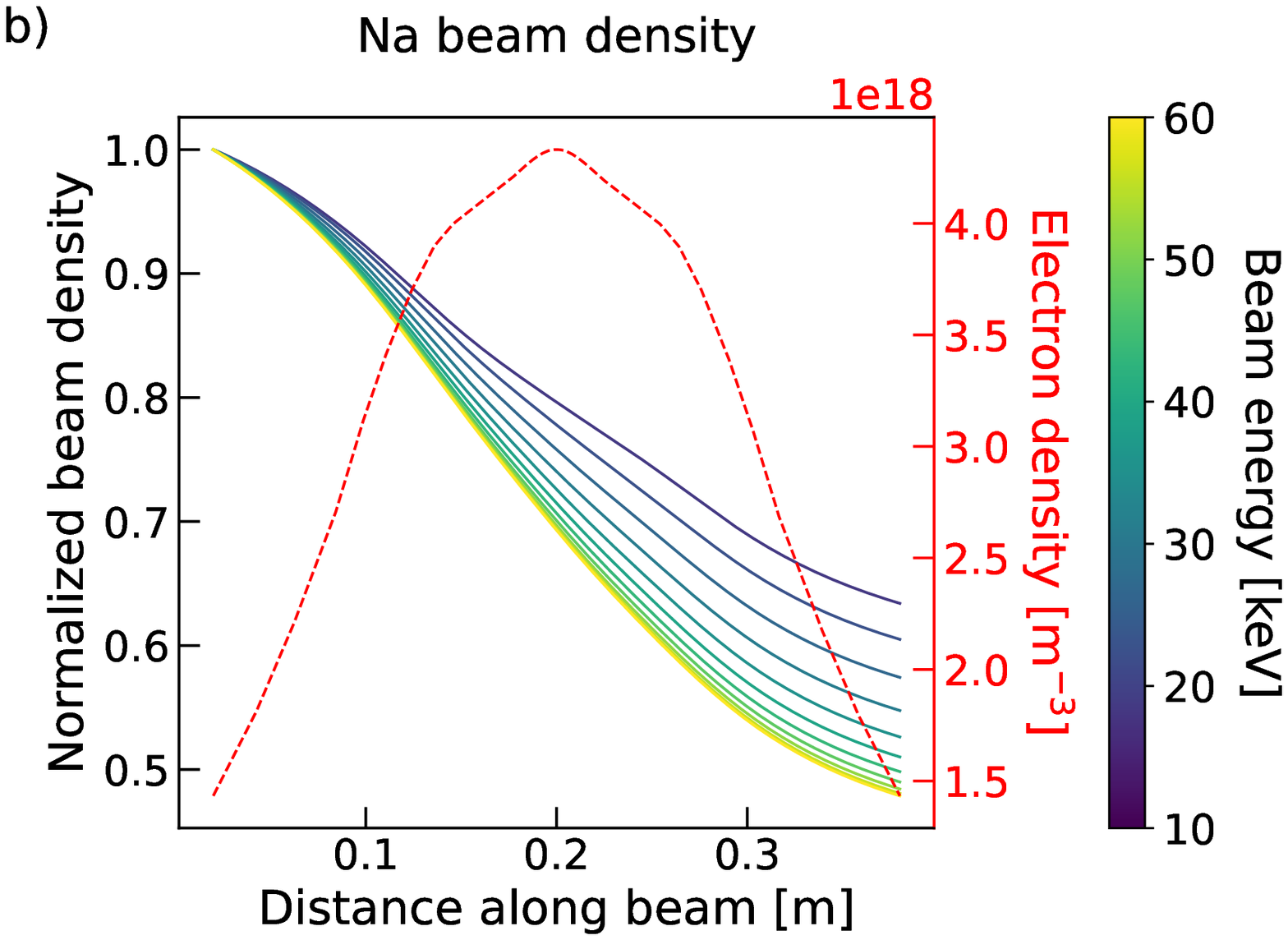}
\caption{The density of lithium (a) and sodium (b) beams in HSX-like plasma, normalized to the initial beam density (solid lines). Beam energy is represented by the coloring of the lines. The electron density of the plasma is also shown with the red, dashed line. An inverse relation between beam energy and penetration depth is observable under 25 keV for lithium, and in the whole energy range for sodium.}
\label{fig:nbd}
\end{center}
\end{figure*}

The evolution of lithium and sodium beams calculated by RENATE\nobreakdash-OD for various energies are plotted in Figure \ref{fig:nbd}. Normalized beam density values are used, which are acquired through normalization by the initial beam density. The beam energies applied are in the range of 10-60 keV, which is the usual operating range of alkali diagnostic beams.

The first detail worth noting is the overall attenuation of the beams across the whole plasma. The beam density at the far end of the simulated region is still 50\%-62\% of the initial density, meaning that a significant portion of the beam atoms would reach the opposite wall from the injection site. Also, in a more usual high-ion-temperature plasma, we expect the penetration depth to increase with beam energy \cite{asztalos2022modell}, which is still the situation for lithium beams above 25 keV acceleration. However, lithium beams under 25 keV and sodium beams in the whole energy range exhibit an inverse relation, meaning that beams with lower energy can penetrate the plasma better.

\begin{figure*}
\begin{center}
\includegraphics[width=.4\linewidth]{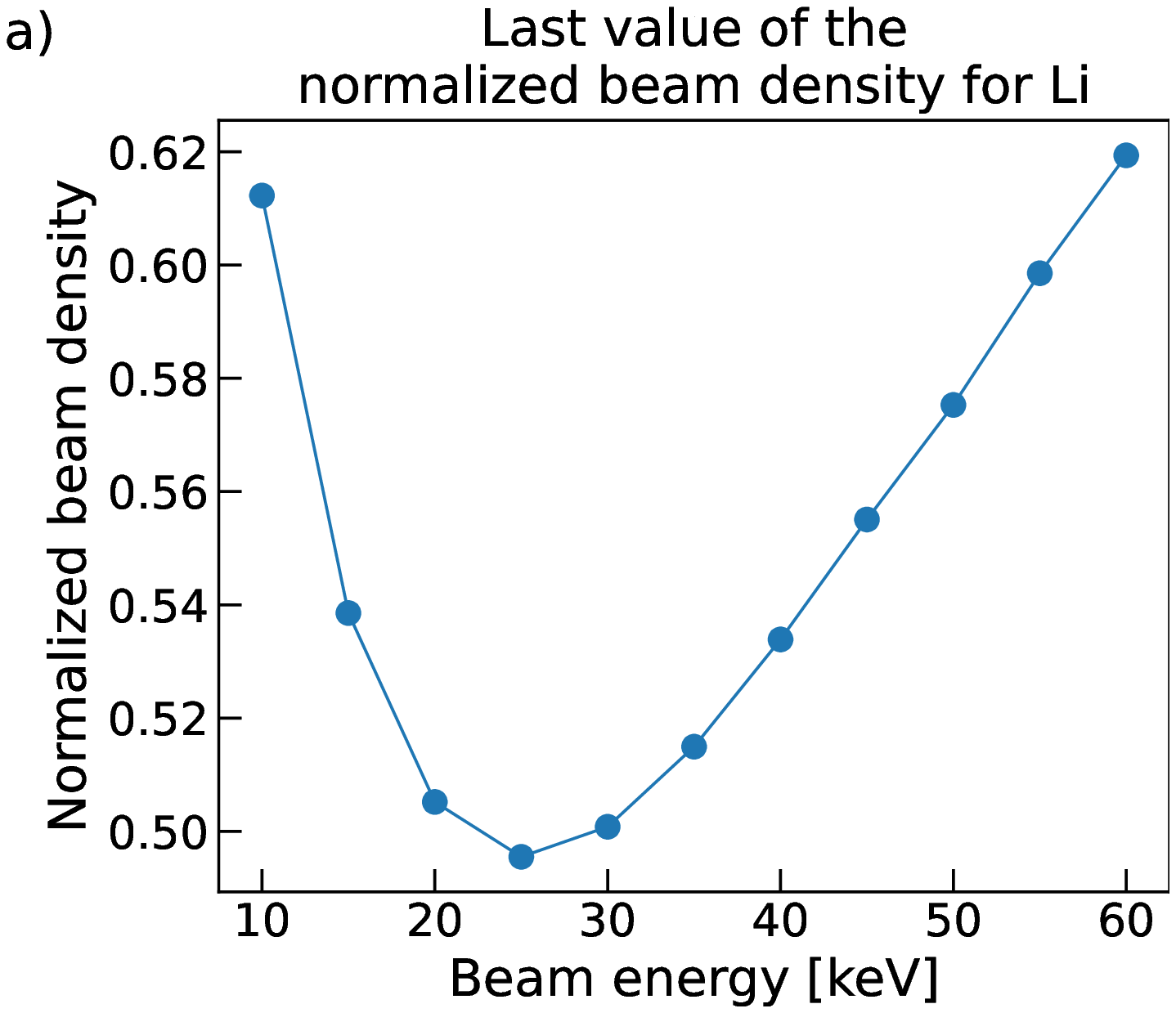}
\includegraphics[width=.4\linewidth]{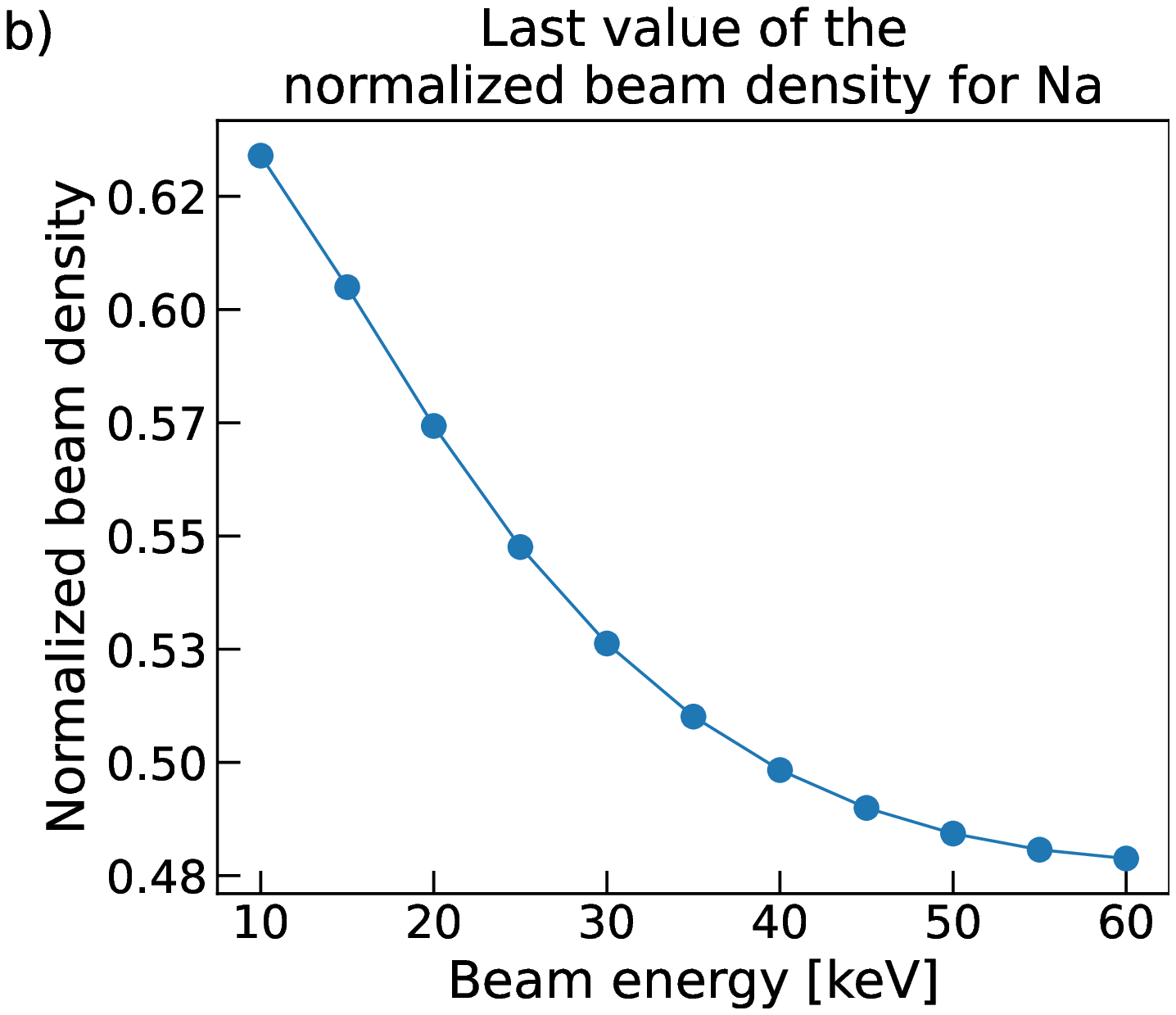}
\caption{Final values of the normalized beam density at the end of the simulated region. Increasing penetration depth with decreasing energy is only present under 25 keV beam energy with lithium (a), while with sodium this is the case for the whole energy range (b).}
\label{fig:last_nbd}
\end{center}
\end{figure*}

In Figure \ref{fig:last_nbd} the last normalized beam density values of the simulated region are plotted against beam energy, showing the remaining portion of the beam atoms after traveling through the plasma. In the case of lithium, the usual tendency of less attenuation with increasing beam energy still holds between 25 and 60 keV, but the situation reverses under 25 keV. For sodium beams, the whole simulated energy range shows the inverse tendency.

This behavior can be better understood by examining the underlying cross-sections and the corresponding reduced rate coefficients. In particular, it is enough to study the electron loss process from the ground state of the beam atoms, since this is the main process behind beam attenuation, and processes of the higher levels usually follow similar trends as those of the ground level. Additionally, the large difference between the temperature of the ion and electron population in the plasma causes the attenuation to be dominated by proton impacts, therefore the analysis is concentrated on this process. The following section describes the corresponding rate coefficients for sodium beams, but the main effects are similar in nature for the lithium beams, too.

\begin{figure}
\begin{center}
\includegraphics[width=\linewidth]{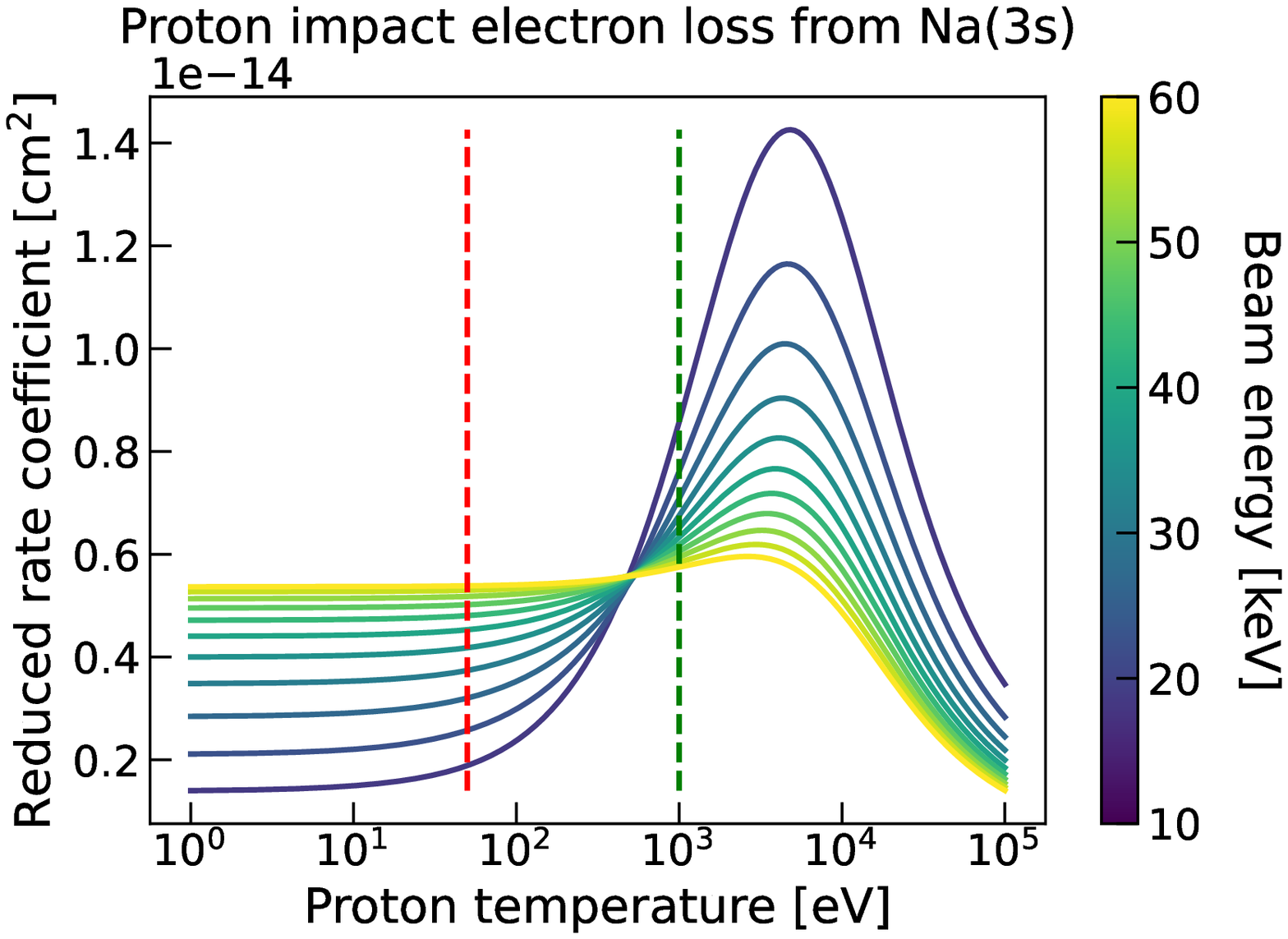}
\caption{Reduced rate coefficients calculated with equations (\ref{equ:rate}) and (\ref{equ:redrate}) for proton impact electron loss from Na(3s) (solid lines). The different beam energies are represented by different colors. The red, dashed line marks 50 eV, the ion temperature in the simulations, while 1 keV was chosen as a generic temperature to represent the region of usual behavior.}
\label{fig:ion_rates}
\end{center}
\end{figure}

The reduced rate coefficients are shown in Figure \ref{fig:ion_rates} as continuous functions of ion temperature for the different beam energies. At the ion temperature of the HSX plasma, marked with the red and dashed line, higher beam energy leads to higher attenuation. However, this relation reverses at a few hundred electronvolts, and higher beam energy leads to lower attenuation above this transition. 1 keV was chosen as a generic example to represent this region.

\begin{figure*}
\begin{center}
\includegraphics[width=.41\linewidth]{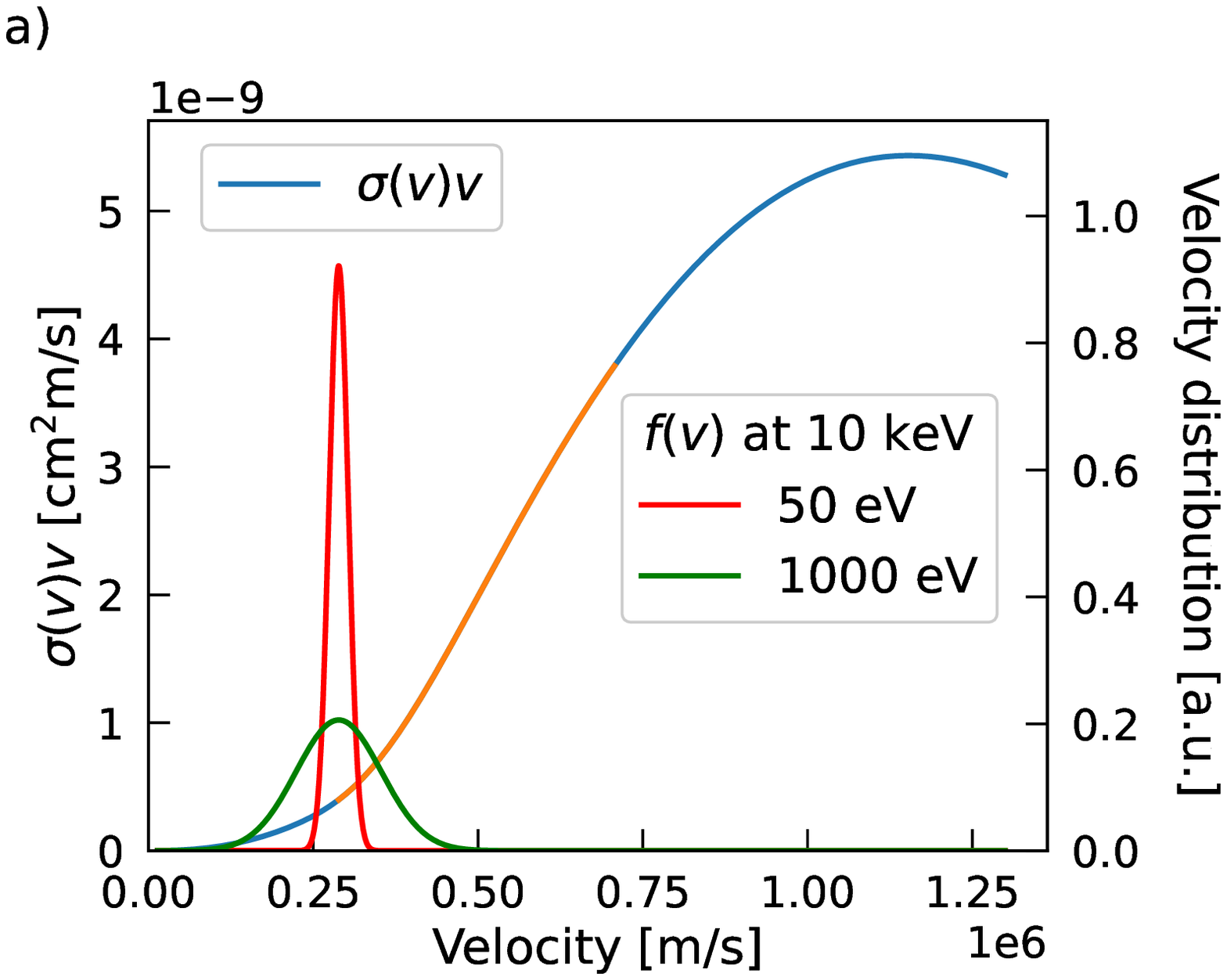}
\includegraphics[width=.41\linewidth]{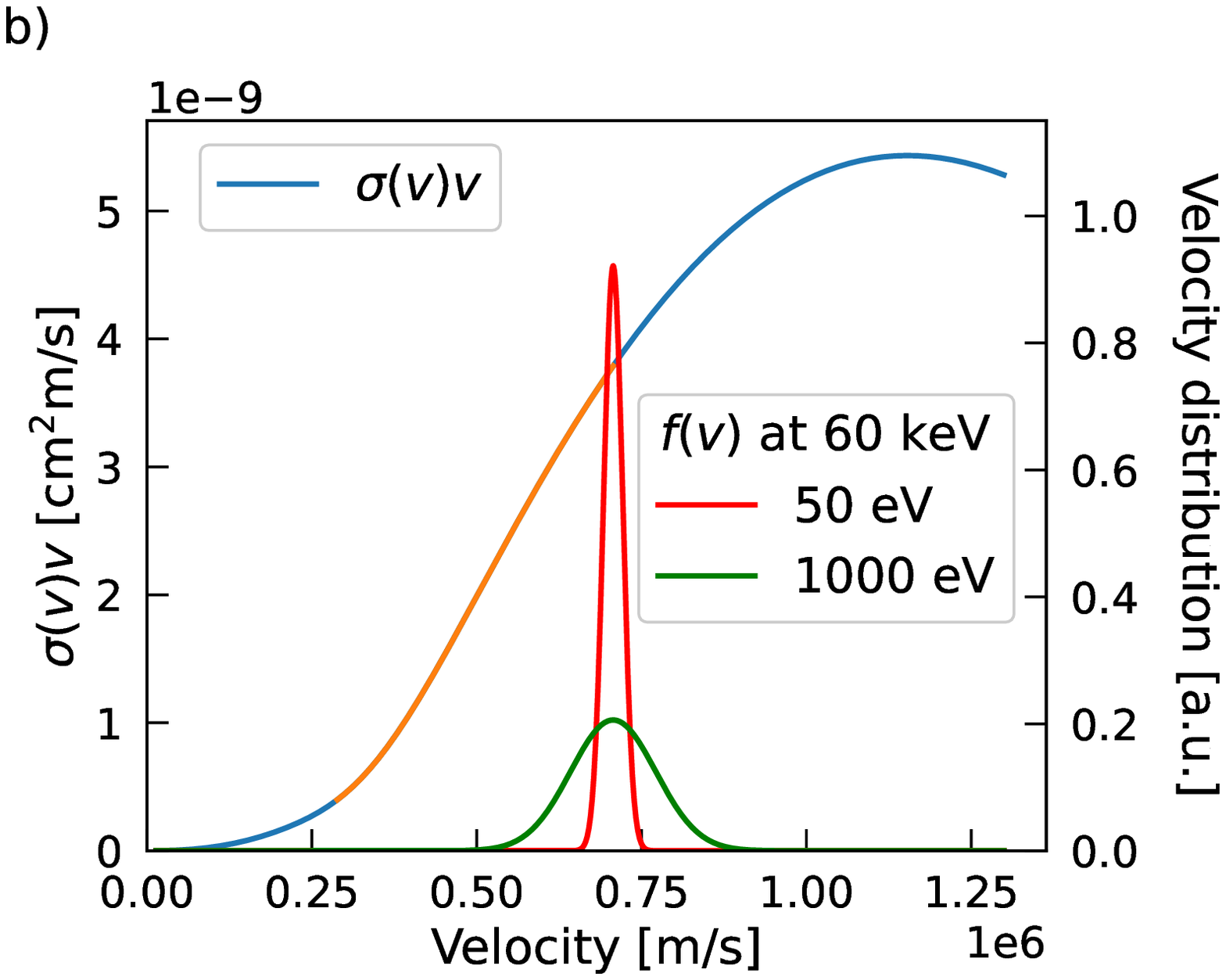}
\caption{a) Cross-section of electron loss from the 3s level of sodium due to proton impacts, the main channel behind beam attenuation, multiplied with the relative velocity as in equation (\ref{equ:rate}). The highlighted section is the velocity range covered by the simulated beam energies. The remaining term under the integral in formula (\ref{equ:rate}), the relative velocity distribution function, is also plotted for both 50 eV and 1 keV plasma temperature for a 10 keV plasma temperature beam. b) The same curves are plotted for a sodium beam with 60 keV energy.}
\label{fig:low_cross}
\end{center}
\end{figure*}

To understand the connection between the reduced rate coefficients and ion temperature, we have to examine the quantities found in the integral of equation (\ref{equ:rate}). The $\sigma(v)v$ product and separately the distribution function $f(v,T)$ for the proton impact electron loss rate from the $3s$ level of 10 keV sodium atoms are shown in Figure \ref{fig:low_cross}/a. For the illustration of how plasma temperature changes the rate coefficients, there are two distribution functions are plotted, one according to 50 eV and one according to 1 keV ion temperature. Note that both of these functions are centered on the same relative velocity determined by the beam energy. When the plasma temperature is low, the distribution function resembles a Dirac delta, so the rate coefficient is relatively low. However, increasing the temperature widens the distribution function, which combined with the convex nature of the $\sigma(v)v$ product increases the rate coefficient.

On the other hand, the situation is different when we perform these calculations for a beam with higher, 60 keV energy. The same curves are plotted in Figure \ref{fig:low_cross}/b for this case. It is immediately visible that the distribution functions are shifted to a higher velocity. This has two effects. First, with 50 eV plasma temperature, the resulting rate coefficient in Figure \ref{fig:ion_rates} is higher for the high energy beam, since the Dirac-delta-like distribution function selects a higher value from the $\sigma(v)v$ product. However, in this velocity range, the curve of this product is concave, so using the wider distribution function of 1 keV plasma protons only slightly increases the rate coefficient.

The temperature dependence of proton impact electron loss rate coefficients and the low ion temperature explain the inverse relation between beam energy and penetration depth in the HSX study. Usually, neutral beams operate in plasma where beam attenuation is driven by both a high-temperature ion and high-temperature electron population, leading to low-energy beams suffering from higher electron loss rates. However, in the low-ion-temperature and high-electron-temperature plasma of HSX, this tendency reverses completely for sodium beams, and partially for lithium beams in the examined energy region, meaning that low-energy beams experience lower electron loss rates.

\subsection{Beam emission}

Understanding the evolution of beam density is an important step when evaluating the simulation results, but the quantity more directly related to the performance of a BES diagnostic system is the emission density of the beam. This is acquired by multiplying the population density of the upper level of the transition we are interested in with the spontaneous transition rate to the lower level. Since we calculated one-dimensional beam evolution, this gives us the linear emission density of the simulated beam. The absolute value of this quantity is still proportional to the beam current, which was chosen to be 1 mA.

\begin{figure}
\begin{center}
\includegraphics[width=\linewidth]{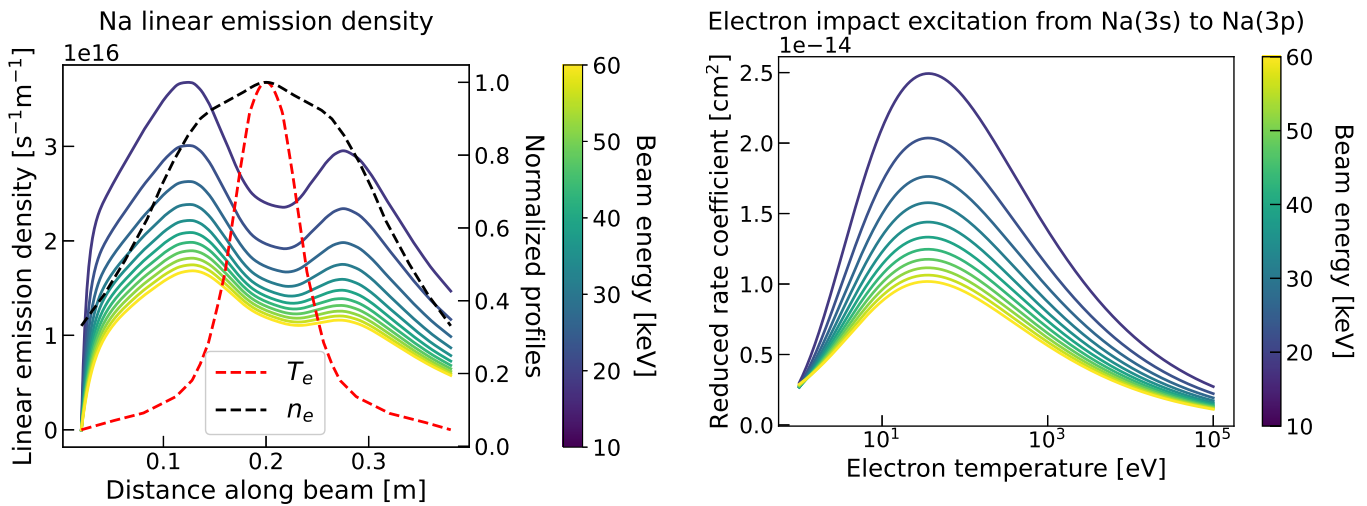}
\caption{Linear emission density of sodium beams with different energy (solid lines). The electron temperature and density profiles are also plotted with red and black dashed lines, respectively. Beams with lower energy emit more light, which is expected due to the lower speed of the beam atoms, giving more chance for excitation over a given distance. However, such a strong reaction to the increase in electron temperature is rarely seen, since usually beam evolution is mostly driven by plasma density. In this case, the high electron temperature in the core leads to a lower rate of excitation by electron impact, which causes the emission density to decrease in the core.}
\label{fig:emission}
\end{center}
\end{figure}

The linear beam emission density is plotted for the previously discussed sodium beams in Figure \ref{fig:emission}. The transition we considered is 3p~$\rightarrow$~3s, as the one observed in existing diagnostic systems \cite{wolfrum2009sodium}. As expected, beams with lower energy provide higher peak emission densities due to their lower velocities through the plasma, providing a higher probability for excitation over a given distance. 

The evolution of the emission density is fairly unusual, with high emission across the whole plasma, and peaks on both sides of the core. The high emission is evidently due to the low plasma density, and therefore low beam attenuation, as seen in the previous section. The performance of a BES system is highly dependent on the SNR of measurements, which is in a trade-off relation with the spatial resolution of the setup. However, regardless of the balance decided between these two properties, a beam with high emission density is always beneficial for BES. The best way to give an estimation of how much of a benefit the low plasma density means for the emission density is to estimate the SNR with some generic parameters. Assuming that a detector integrates $2\cdot10^{16}$~s$^{-1}$m$^{-1}$ emission over a distance of 1~cm, we get $2\cdot10^{14}$~s$^{-1}$ photon current in total, which would reduce to around $2\cdot10^{11}$~s$^{-1}$ collected photon current after passing through the observation system. Depending on the detector type, this means an SNR of 30-160 when sampling with 1 MHz \cite{dunai2010avalanche}. As a comparison, typical SNR values are around 5 in JET \cite{refy2018sub}, 10 in ASDEX-U \cite{Willensdorfer_PPCF_2014}, while 20-50 in EAST \cite{Zoletnik_EAST_RSI_2018}, KSTAR \cite{lampert2015combined} and W7-X \cite{Zoletnik_RSI_ABES_2018}.

\begin{figure}
\begin{center}
\includegraphics[width=\linewidth]{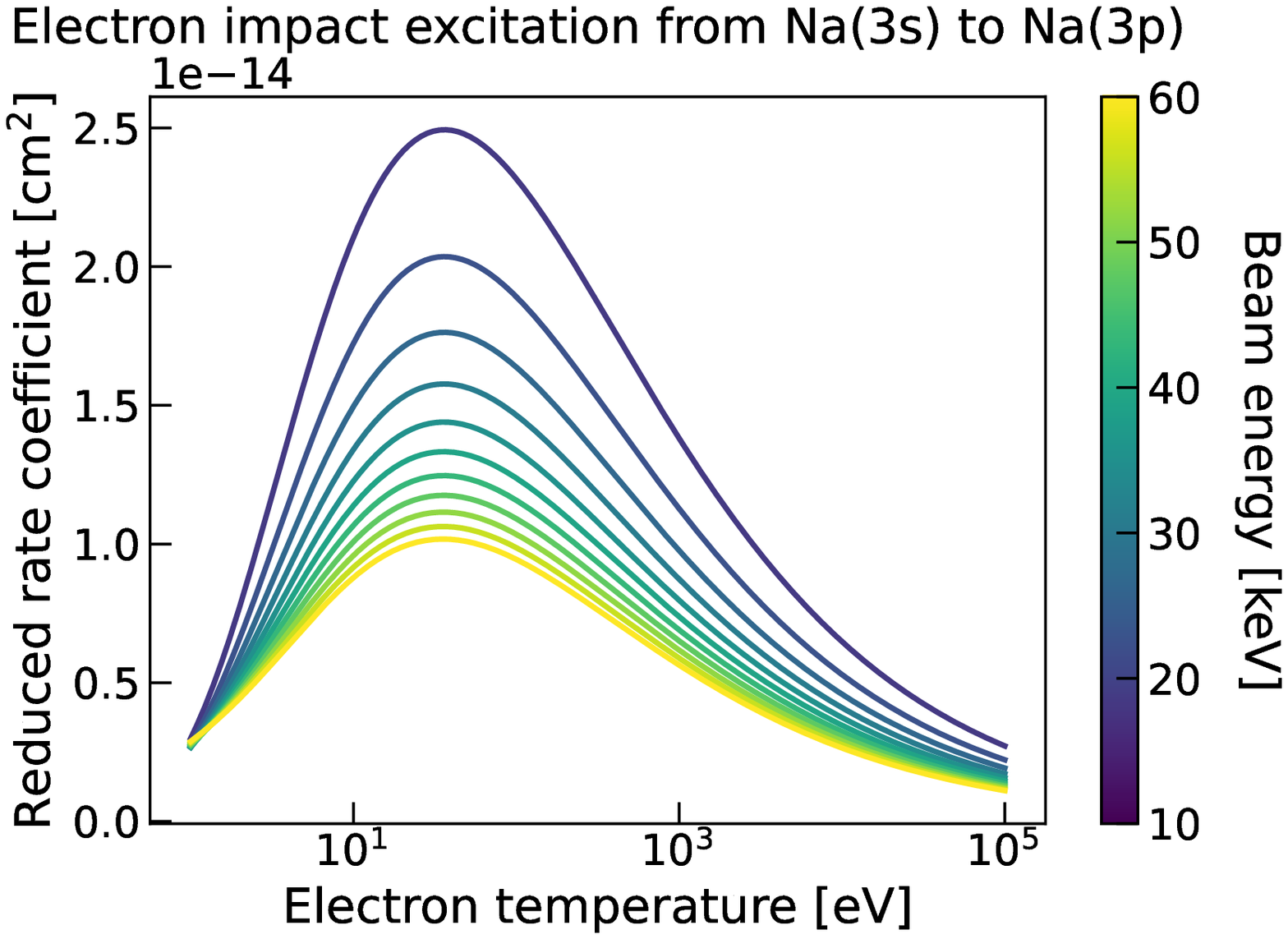}
\caption{Reduced rate coefficients in the function of electron temperature for electron impact induced excitation from Na(3s) to Na(3p), the main reaction channel populating the 3p level. In the temperature range of $\sim$100-2500 eV present in the HSX plasma, the rate coefficients are decreasing for all beam energies.}
\label{fig:ex_rates}
\end{center}
\end{figure}

Regarding the two distinct emission peaks, the explanation lies within the temperature dependence of rate coefficients governing the population of the 3$p$ level. In particular, the most dominant reaction populating this level is electron impact excitation from the 3$s$ level. The reduced rate coefficients for this process are plotted in Figure \ref{fig:ex_rates} for different beam energies against electron temperature. In the temperature region of the HSX plasma profile ($\sim$100-2500~eV), the rate coefficients are steadily decreasing with increasing temperature, which explains the behavior seen in the emission profile. As the beam is traveling through the plasma, first the emission is increasing as the higher atomic levels get populated due to the increasing plasma density. However, as the beam reaches the core plasma, there is a rapid increase in electron temperature, leading to a significant decrease in the rate at which the 3$p$ level gets populated. This is mirrored by a decrease in the emission as well. After traveling through the core plasma, the electron temperature drops again, leading to an increase in the beam emission.

Of course, this effect is always present, but the plasma in HSX is unique due to its low density and low beam attenuation, allowing the emission to maintain a high intensity throughout the plasma, and also due to its extremely peaked electron temperature profile, making a strong impact on the emission profile.

\section{Summary and outlook}
\label{sum}

Recently, we completed a feasibility study for the HSX stellarator, which has the unique trait among fusion experiments of studying low-ion-temperature plasma configurations. In the study, we examined the expected performance of an alkali beam emission diagnostic system by simulating the beam evolution with our in-house code, RENATE\nobreakdash-OD. We performed simulations for both lithium and sodium beams in the energy range of 10-60 keV, propagating through a plasma profile acquired from earlier experiments.

The results, namely the relation between plasma penetration and beam energy, showed unusual effects compared to other BES diagnostics working in high-ion-temperature plasma. Instead of the penetration increasing with beam energy in the whole energy region, we found that low-energy beams achieve better penetration. The explanation for this effect is in the underlying electron loss cross-sections and the corresponding rate coefficients. The temperature-dependent tendencies of the rate coefficients are determined by the curvature of the $\sigma(v)v$ product in the relevant relative velocity region during the integration with the ion distribution function. The center of the relevant region is at the velocity of beam atoms, and its width is determined by the temperature of the ion population, so both beam energy and ion temperature may have a significant effect on the rate coefficients. In case of the HSX-like conditions, this relation manifests in low-energy beams experiencing lower attenuation, than the high-energy beams of the study.

Apart from beam attenuation, we also examined the beam emission density, as an important quantity for the capabilities of a BES diagnostic system. The unusual plasma parameters manifested some interesting effects in this case as well. First, the low plasma density allows the emission density to remain high throughout the plasma, unlike in high-density plasma configurations, where the emission reaches its peak and decays in a few centimeters. This also allows the peaked electron temperature profile to make its impact on the beam emission, since the high temperature in the core lowers the rate of the exited states getting populated, resulting in decreased emission density in the core.

Altogether, it is clear that the diagnostics of low-ion-temperature and low-density plasma with beam emission spectroscopy feature effects unseen in high-density experiments. While the former conditions are far from those of a fusion-supporting plasma, the outer regions and the divertor vicinity of future large-scale devices, equipped with more exotic divertor configurations like the Super-X in MAST-U, are expected to hold plasma resembling the HSX conditions \cite{havlivckova2013numerical, havlivckova2014investigation, havlivckova2015solps}. These regions form the interface between the plasma and the chamber, so monitoring them is important for both scientific and operational purposes. If BES is ever used for the diagnostics of these areas, the effects described in our study should be considered carefully.

\section*{Acknowledgments}
Supported by the KDP-2021 Program of the Ministry for Innovation and Technology from the source of the National Research, Development and Innovation Fund.

G.I. Pokol, P. Balázs and O. Asztalos acknowledge the support of the National Research, Development and Innovation Office (NKFIH) Grant FK132134.

This work has been carried out within the framework of the EUROfusion Consortium, funded by the European Union via the Euratom Research and Training Programme (Grant Agreement No 101052200 - EUROfusion). Views and opinions expressed are however those of the author(s) only and do not necessarily reflect those of the European Union or the European Commission. Neither the European Union nor the European Commission can be held responsible for them.

\bibliography{p_balazs_soft_2022_arxiv}

\end{document}